\documentclass[%
 reprint,
 amsmath,amssymb,
 aps,
 prl
]{revtex4-2}

\usepackage{graphicx}
\usepackage{dcolumn}
\usepackage{bm}
\usepackage{todonotes} 
\usepackage{hyperref}
\usepackage[mathlines]{lineno}
\usepackage{multirow}
\usepackage{adjustbox}
\usepackage{units}

\newcommand{\edit}[1]{{\color{black}{#1}}}

\begin{document}
\preprint{APS/123-QED}


\title{High brightness multi-MeV photon source\\ driven by a petawatt-scale laser wakefield accelerator}
\newcommand{\IC}{The John Adams Institute for Accelerator Science, Imperial College London, London, SW7 2AZ, UK}

\newcommand{\UMICH}{G\'{e}rard Mourou Center for Ultrafast Optical Science, University of Michigan, Ann Arbor, MI 48109-2099, USA}

\newcommand{\QUB}{School of Mathematics and Physics, Queen's University of Belfast, BT7 1NN, Belfast, UK}

\newcommand{\CLF}{Central Laser Facility, STFC Rutherford Appleton Laboratory, Didcot OX11 0QX, UK}

\newcommand{\YORK}{Department of Physics, University of York, York, YO10 5DD, UK}

\newcommand{\ELINP}{Extreme Light Infrastructure - Nuclear Physics, ``Horia Hulubei'' National Institute for R\&D in Physics and Nuclear Engineering, 30 Reactorului Street, 077125 Magurele, Romania}

\newcommand{\PRINCETON}{Department of Astrophysical Sciences, Princeton University, Princeton, New Jersey 08544, USA.}

\newcommand{\qub}{$^1$}
\newcommand{\imp}{$^2$}
\newcommand{\elinp}{$^3$} 
\newcommand{\clf}{$^4$}
\newcommand{\umich}{$^5$} 
\newcommand{\york}{$^6$}
\newcommand{\princeton}{$^7$}

\def\andname{,}
\def\andname{\hspace*{-0.5em},}

\author{
E.~Gerstmayr\qub$^{,}$}
\email{e.gerstmayr@qub.ac.uk}
\author{
B.~Kettle\imp,
M.J.V.~Streeter\qub,
L.~Tudor\elinp,
O.J.~Finlay\clf,
L.E.~Bradley\imp,
R.~Fitzgarrald\umich,
T.~Foster\qub,
P.~Gellersen\york,
A.E.~Gunn\imp,
O.~Lawrence\york,
P.P.~Rajeev\clf,
B.K.~Russell\princeton,
D.R.~Symes\clf,
C.D.~Murphy\york,
A.G.R.~Thomas\umich,
C.P.~Ridgers\york,
G.~Sarri\qub,
~and~
S.P.D.~Mangles\imp
}




\address{\qub \QUB} 
\address{\imp \IC} 
\address{\elinp \ELINP} 
\address{\clf \CLF} 
\address{\umich \UMICH}  
\address{\york \YORK} 
\address{\princeton \PRINCETON}

\vspace{10pt}

\date{\today}


\begin{abstract}
We present an experimental demonstration of a bright multi-MeV gamma source driven by a petawatt laser.
The source generates on average $\unit[(1.2\pm0.6)\times10^9]{photons}$ above $\unit[1]{MeV}$ per pulse, exceeding those of previous all-optical sources by a hundred times, and reached a peak spectral brightness of $\unit[(3.9 \pm 1.5)\times 10^{22}]{photons/mm^2/mrad^2/s/0.1\%BW}$ at $\unit[\epsilon_\gamma\approx11]{MeV}$.
The source was produced by inverse Compton scattering of a laser wakefield accelerated GeV electron beam and its back-reflected driving laser pulse.
\edit{Its performance is well described by a simple model of the laser and electron properties at the collision point
that allows quantitative predictions and identifies clear strategies to further enhance radiation efficiency.}
Our results highlight the promise of this source for fundamental physics studies, as well as for applications of nuclear resonance fluorescence and nuclear transmutation.

\end{abstract}


\maketitle


Particle accelerator based X-ray sources have a long history as critical tools in scientific discovery. 
Synchrotrons and X-ray free electron lasers (XFELs) \cite{EmmaNatPhoton2010,IshikawaNatPhoton2012,DeckingNatPhoton2020} provide high flux and, in the case of XFELs, extremely bright sources of radiation in the \unit[$\epsilon_\gamma\lesssim 100$]{keV} range for a wide range of applications in medicine, biology, chemistry, fundamental physics fluorescence studies and industry. 
Due to practical limitations of magnetic undulators, achieving energies significantly beyond \unit[$100$]{keV} currently requires different processes such as bremsstrahlung  \cite{SoberNIMA2000,KaiserNIMA2008} and inverse Compton scattering (ICS) \cite{DuRSI2013,OhgakiNIMA2000,NiknejadiPRAB2019,GuntherSynchrotronRad2020,WangNST2022,SandorfiIEEE1983,MuramatsuNIM2022}.
Of these two processes, ICS can generate higher spectral intensities at high photon energy.

In ICS, a laser pulse scatters off a relativistic electron, boosting the photon energy to $\epsilon_\gamma \simeq 4 \gamma^2 \epsilon_i$, where $\epsilon_i$ is the energy of the incident laser photon, and $\gamma$ is the relativistic Lorentz factor. 
In the case of ultrarelativistic electrons ($\gamma\sim10^3$) and optical photons (\unit[$\epsilon_i\sim1$]{eV}) this results in the emission of multi-MeV radiation.
Several ICS facilities are operational worldwide, typically combining a storage ring or linear accelerator with a laser cavity to generate photon sources in the sub-MeV range. 
Only a few large-scale facilities are capable of accessing \unit[$\epsilon_\gamma>1$]{MeV} energies, e.g. TERAS LCS \cite{OhgakiNIMA2000}, HI$\gamma$S \cite{WellerNPN2015}, SLEGS at SSRF \cite{WangNST2022} and SPring-8 LEPS2 \cite{MuramatsuNIM2022}, due to the requirements for high electron energies.
In this range, ICS sources have been used for nuclear resonance fluorescence \cite{BertozziNIMPB2005,KikuzawaAPE2009},  the study of photonuclear reactions \cite{HaradaPRL1998,Quiter2012PhysRevC,Zilges2022PPNP}, nuclear waste treatment \cite{ChenCPC2008,LiJNT2009,RehmanANE2017}, isotope production \cite{HabsAppPhysB2011}, and particle physics applications through hadron photoproduction \cite{MuramatsuNIM2022}.

Laser-wakefield acceleration (LWFA), offers an attractive alternative for the \unit[$\epsilon_\gamma>$]{MeV} photon energy range, due to its ability to obtain GeV-scale electron energies in centimetre-scale plasmas \cite{LeemansNatPhys2006,KneipPRL2009} as well as their convenient co-location with high-power laser systems.
In LWFA, a high-intensity laser pulse drives a plasma wave to accelerate an electron beam, which subsequently undergoes scattering with a second laser pulse.
Several experiments have demonstrated photon beams with energies between \unit[100]{keV} and \unit[1]{GeV} \cite{ChenPRL2013,PowersNatPhoton2014,SarriPRL2014,KhrennikovPRL2015,YanNatPhoton2017,MirzaieNatPhoton2024} 
and can generate unprecedented peak spectral brightness \cite{MirzaieNatPhoton2024}, due to their ultra-short pulse durations (\unit[$\sim$]{fs}) \cite{LundhNatPhys2011} and small source size (\unit[$\sim$]{{\textmu}m}) \cite{WeingartnerPRAB2012}.

\begin{figure*}
    \centering
    \includegraphics[width=0.99\linewidth]{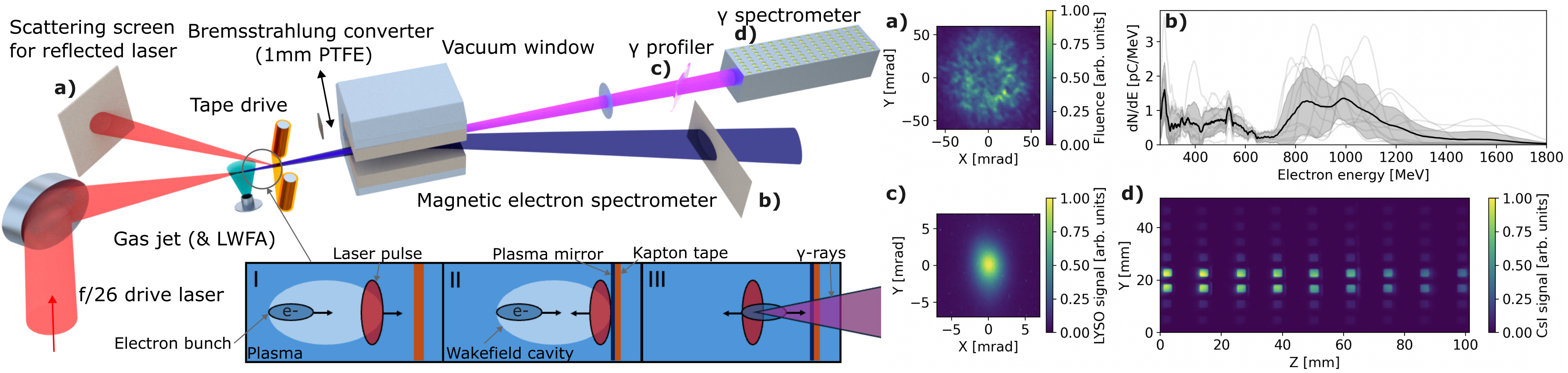} 
    \caption{\textbf{Left:} Sketch of the experimental setup with key diagnostics. \textbf{Right:} Example diagnostic images: \textbf{(a)} scattering screen, \textbf{(b)} electron spectra (grey lines) and their average spectrum (standard deviation from mean in shaded), \textbf{(c)} gamma profiler and \textbf{(d)} spectrometer. \textbf{Inset:} \textbf{(I)} The laser pulse (red) drives a wake and accelerates an ultrarelativistic electron ($e^-$) bunch (blue). \textbf{(II)} The laser pulse ionises the surface of the kapton tape (orange) and forms a plasma mirror that reflects it. \textbf{(III)} The reflected laser pulse collides with the electron beam and emits radiation from inverse Compton scattering (purple).}
    \label{fig:setup}
\end{figure*}

All-optical ICS can also be produced with a single laser pulse, by using a plasma mirror to reflect the LWFA laser driver back onto the accelerated electron beam (see inset of Fig. \ref{fig:setup}), as demonstrated by \textit{Ta Phuoc et al.} in 2012 \cite{TaPhuocNatPhoton2012}.
This `self-aligned' scheme guarantees alignment on every shot (a challenge for two-beam experiments \cite{SamarinJMO2018,KimRSI2022}) and essentially halves the laser requirements while enabling ICS sources at the more commonly existing single-beam laser facilities. 
\edit{For practical, high-impact applications, it is critical to optimise the ICS efficiency to maximise photon yield with limited laser energy.}
Initial experiments have demonstrated this scheme at laser powers of \unit[30]{TW} yielding up to 100 keV photon energies \cite{TaPhuocNatPhoton2012,TsaiPoP2015,DoppPPCF2016}, extending up to MeV photon energies at the \unit[100]{TW} scale \cite{YuSciRep2016,WuPPCF2019}.
With this work, we extend the single-beam approach to the recently more widely available petawatt laser powers.
\edit{By generating photons with energies exceeding \unit[10]{MeV}, the source becomes directly relevant to nuclear physics applications and substantially broadens the range of potential uses.}
Its high peak flux and ultrafast pulse duration provide novel capabilities that complement existing radiofrequency-based ICS sources, e.g. enabling imaging of fast processes or studies of fundamental physics such as photon-photon scattering \cite{DrebotPRAB2017}.



The experiment was conducted at the Extreme Light Infrastructure - Nuclear Physics, Romania, in the \unit[1]{PW} Experimental Area - E5 \cite{ELINPHPLSE2020}.
An overview of the experimental setup is shown in Figure \ref{fig:setup}.
The drive laser (\unit[810]{nm}) was focused using an $f/26$ parabolic mirror to a spot size of \unit[32]{\textmu m} \textsc{fwhm}, with the central \textsc{fwhm} spot containing \unit[14]{\%} of the energy.
Each laser pulse delivered $\unit[20.0 \pm 0.3]{J}$ energy on-target at a pulse duration of \unit[$\tau\gtrsim25$]{fs} and linear polarisation in the horizontal plane, reaching a peak intensity of \unit[$I_0=(1.6\pm0.2)\times 10^{19}$]{W/cm$^2$} (normalised vector potential $a_0 = 2.6 \pm 0.2$). 
The laser pulses were focused into a 20-mm supersonic gas jet, \unit[30]{mm} above the nozzle, and drove a LWFA in a plasma of density of $n_e \approx \unit[1.75\times10^{18}]{cm^{-3}}$ containing \unit[98]{\%} helium and \unit[2]{\%} nitrogen as a dopant.
The repetition rate of the \unit[1]{Hz} laser system was limited to \unit[0.025]{Hz} to reduce the gas load on the vacuum system.
The LWFA generated GeV-scale electron beams were measured using a magnetic spectrometer (with a low-energy cut-off at \unit[260]{MeV}), consisting of a permanent dipole magnet with $\int B \mathrm{d}x = \unit[0.3]{T m}$ 
and charge-calibrated scintillating Lanex screens \cite{FreemanRSI2011,WilliamsRSI2014}.

At the exit of the accelerator, a replenishable tape (\unit[25]{\textmu m} Kapton) was inserted, tilted by 4 degrees, to act as a plasma mirror. 
This reflected the residual laser onto a calibrated scattering screen, providing a measurement of plasma mirror reflectivity as $\unit[\approx40]{\%}$. 
The laser pulse was closely followed by the accelerated electron beam, resulting in gamma emission via ICS once the laser was reflected (see inset of Fig. \ref{fig:setup}).
The pointing stability of the reflected laser (standard deviation from the mean position) was $\unit[2]{mrad}\times\unit[6]{mrad}$, increasing to at most $\unit[14]{mrad}\times\unit[39]{mrad}$ when the tape was \unit[2.5]{mm} from the jet and perturbed by its gas flow.

The ICS photon beam was emitted collinear with the electron beam axis and propagated through a \unit[2.1]{mm} thick aluminium vacuum window to a suite of gamma diagnostics outside of the vacuum chamber.
The transverse profile was measured using a cerium-doped LYSO (Lu$_2$SiO$_5$:Ce) scintillator with dimensions $\unit[50]{mm}\times\unit[50]{mm}$ and $\unit[2]{mm}$ thickness, placed at 45 degrees relative to the beam axis.
The spectrum was measured using a pixelated scintillator-based gamma spectrometer consisting of thallium-doped caesium-iodide (CsI:Tl) crystals \cite{BehmRSI2018}.
The gamma diagnostics were calibrated by inserting \unit[1]{mm} of PTFE into the electron beam path to generate bremsstrahlung (see Fig. \ref{fig:setup}).
The energy deposited in the detectors was compared with Geant4 modeling to obtain an absolute calibration of the detectors.


The electron beam had a broadband spectrum (see Fig. \ref{fig:setup}b) with a mean energy of 
$\unit[900 \pm 30]{MeV}$ 
and a maximum energy (95\% percentile) of $\unit[1400 \pm 60]{MeV}$ (standard errors).
The beam had an average charge of $\unit[0.9\pm0.4]{nC}$ above \unit[260]{MeV} (standard and systematic error) 
and a $\unit[3.9 \pm 0.4]{mrad}$ (standard error) 
divergence, measured in the non-dispersion plane.



\begin{figure} 
    \centering
    \includegraphics[width=0.99\linewidth]{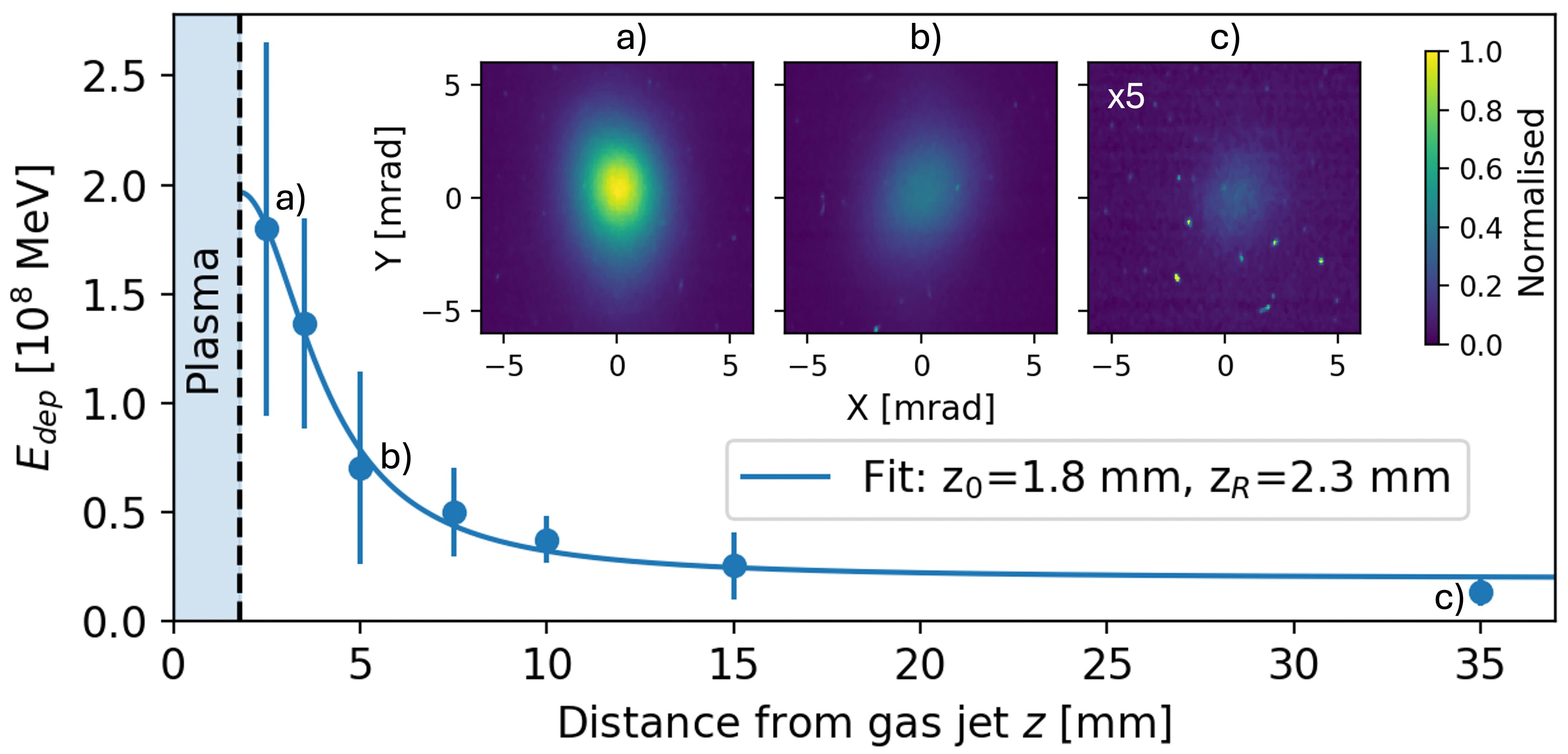} 
    \caption{Average energy deposited in the gamma profiler $E_{dep}$ (error bars for standard deviation) for increasing distance $z$ between the tape and the end of the gas jet.
    The emitted power and $E_{dep}$ are expected to scale with the intensity at the interaction point, which decreases with expanding beam area, 
    i.e. $E_{dep}(z) = A [w_0/w(z)]^2 + C$, where the Gaussian waist is $w(z) = w_0 \sqrt{1+[(z-z_0)/z_R]^2}$. 
    The fitting parameters are the amplitude $A$, the constant offset $C$, the waist position $z_0$, and the Rayleigh length $z_R$.
    The fitting parameters for $z_0$ and $z_R$ are shown in the legend. 
    The dashed line indicates $z_0$.
    \textbf{Inset:} Example gamma profiles at \unit[2.5]{mm} \textbf{(a)}, \unit[5]{mm} \textbf{(b)} and \unit[35]{mm} \textbf{(c)} distance, shown on the same scale. 
    Counts on (c) were multiplied by 5 for visibility. 
    }
    \label{fig:ICSYield}
\end{figure}

The distance from the tape to the exit of the gas jet nozzle was varied from \unit[35]{mm} to \unit[2.5]{mm}.
To avoid damage to the laser system, the tape could not be placed closer than \unit[2.5]{mm} as the gas affected the pointing stability too severely.
The measured energy deposited in the gamma profiler $E_{dep}$ decreased with increasing distance between the tape and the gas jet (see Fig. \ref{fig:ICSYield}), consistent with ICS as the laser pulse diffracts with distance, reducing its intensity at the interaction point.
Other sources of radiation relevant in this context, e.g., bremsstrahlung or betatron radiation (resulting from oscillations of the electron beam in the plasma), do not depend on the position of the tape.
The background signal, e.g. caused by betatron radiation or bremsstrahlung from electrons striking the vacuum chamber walls, was characterised without the plasma mirror and was found to be comparable to the signal for $\unit[z>15]{mm}$ (\unit[$\sim (4.1 \pm 0.1)\times10^7$]{MeV}).
Based on Geant4 simulations, the expected energy deposition in the gamma profiler from bremsstrahlung due to interaction of the electron beam with the tape was found to be negligible ($\sim\unit[(3.3\pm0.6)\times10^6]{MeV}$). 
The signal was dominated by ICS for the closest distances ($\unit[z<5]{mm}$).

\begin{figure} 
    \centering
    \includegraphics[width=0.99\linewidth]{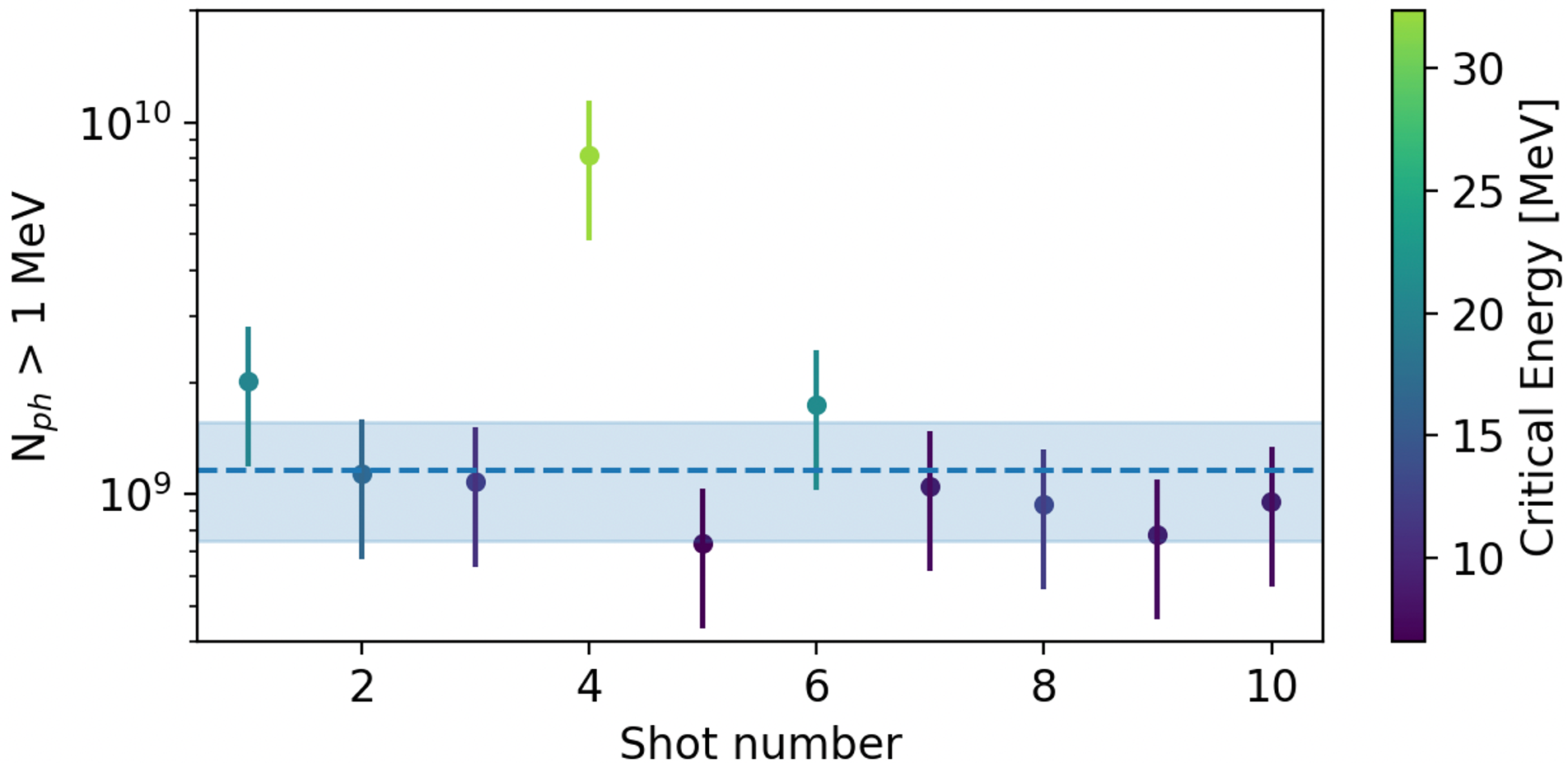} 
    \caption{
    Measured photon numbers above \unit[1]{MeV} (systematic errors as errorbars) for 10 consecutive shots at the closest tape position along with their mean (dashed line) and standard deviation (shaded region), excluding the brightest shot (shot 4). 
    The fitted critical energy is indicated by the marker colour. 
    }
    \label{fig:ICSClose}
\end{figure}

The divergence of the photon beam at the closest tape position was $\unit[(4.0\pm0.5)]{mrad}\times \unit[(3.9\pm0.4)]{mrad}$ \textsc{fwhm}, consistent with the electron divergence, with a pointing variation of $\unit[1.3]{mrad}\times\unit[3.7]{mrad}$ (standard deviation from mean position).
The emitted ICS spectrum is expected to be broadband due to several factors: the large energy spread and divergence of the electron beam, the bandwidth of the laser, and the non-linearity of the laser-electron interaction.
The spectral intensity is described by
$\mathrm{d}I_{\gamma}/\mathrm{d}\epsilon_\gamma = W_{ICS} S(\epsilon_\gamma)$, where $W_{ICS}$ is the radiated energy and the function $S(\epsilon_\gamma)$ is the spectral shape, with $\int_0^\infty S(\epsilon_\gamma)\mathrm{d}\epsilon_\gamma = 1$.
We parametrise $S(\epsilon_\gamma) =[\epsilon_\gamma^{1/3} e^{-\epsilon_\gamma/\epsilon_{crit}}]/[\epsilon_{crit}^{4/3} \Gamma(4/3)]$, which was in simulations found to be a good approximation of the gamma spectrum across a wide range of conditions \cite{ColePRX2018}.
We treat $\epsilon_{crit}$ as a free parameter to control the spectral shape to match the experimental signal \cite{BehmRSI2018}.
In this parametrisation, the mean energy of the emitted photon spectrum is $\epsilon_{crit}/3$.
The measured mean critical energy was $\unit[(12.5\pm1.8)]{MeV}$ (standard error), excluding the exceptionally bright shot at $\unit[\epsilon_\gamma\approx32]{MeV}$ (see Fig. \ref{fig:ICSClose}). 

Using the fitted critical energies and the absolute calibration of the deposited energy, we can infer the number of photons for these shots.
The mean photon number, excluding the brightest shot, was $\unit[(1.2\pm0.6)\times10^9]{photons}$ 
(standard error and systematic error). 
The brightest shot had $\unit[(8.1\pm3.3)\times10^{9}]{photons}$ (systematic error) above \unit[1]{MeV} and the spectrum had a peak of \unit[$\sim10^6$]{photons/0.1\%BW} at \unit[10]{MeV}, which exceeds the photon yields reported in previous all-optical Compton experiments by a hundred times \cite{SarriPRL2014,MirzaieNatPhoton2024}.

While collisions are observed reliably on every shot due to the intrinsic self-alignment, the properties of the photon beam (critical energy, photon numbers) vary from shot to shot. 
This is due to variations in the electron beam (electron energy, charge) and the laser properties at the interaction, caused by shot-to-shot fluctuations in the incident laser pulse and plasma density profile.
The accelerator performance and the laser properties at the interaction are expected to be correlated, e.g. efficient driving of a wakefield resulting in high charge and energy electron beams implies good laser guiding which keeps the laser more focused and intense for longer \cite{StreeterPRAB2022}.

\begin{figure} 
    \centering
    \includegraphics[width=0.99\linewidth]{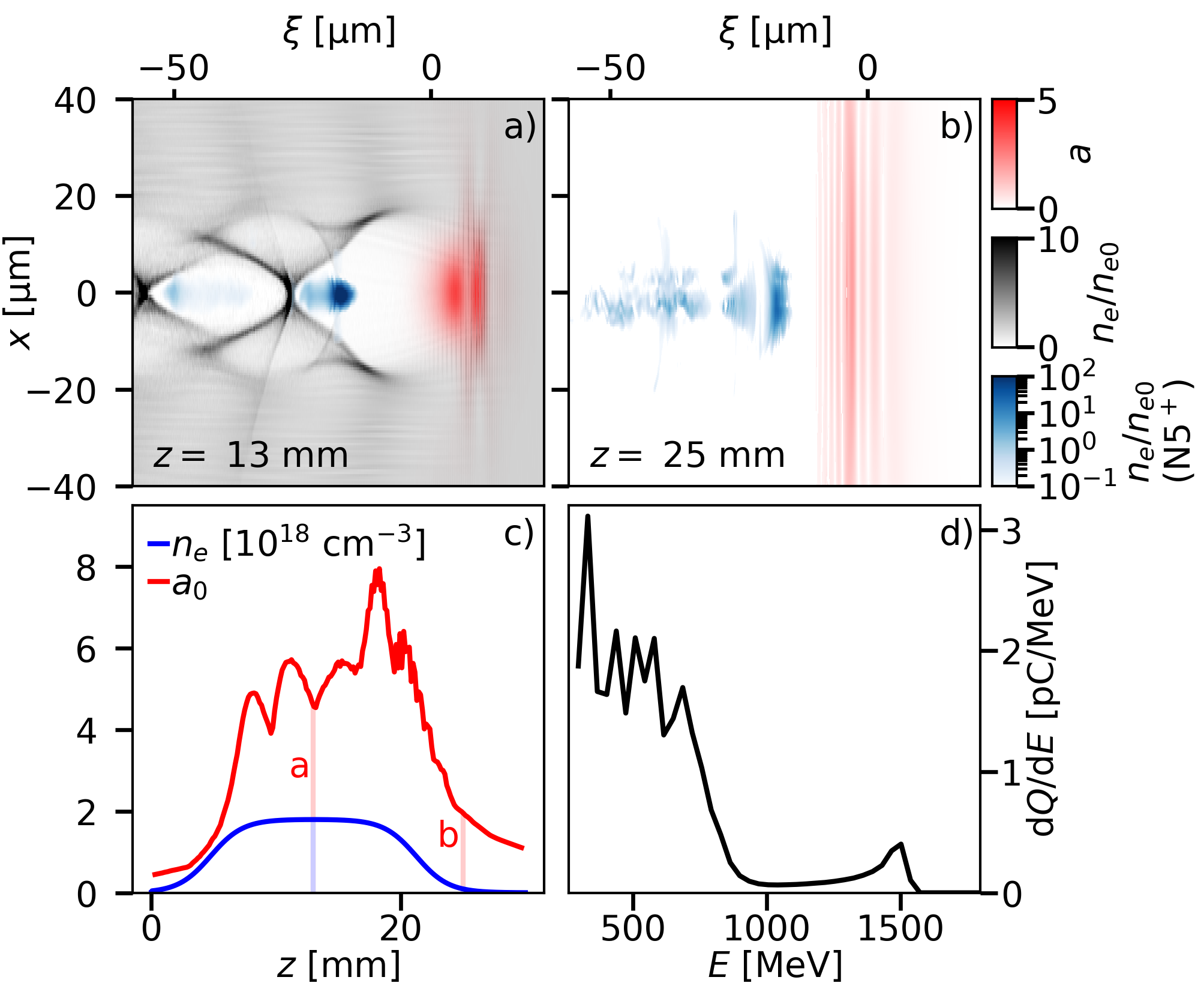}
    \caption{
    FBPIC simulation of the LWFA.
    a) and b) show the laser field, plasma electrons and trapped electron beam at the midpoint and exit of the accelerator respectively, where $\xi = z -ct$ and \unit[$n_{e0}=1.75\times10^{18}$]{cm$^{-3}$} is the electron density at the plateau.
    At the exit plane (corresponding to the closest mirror position in Fig. \ref{fig:ICSYield}) the laser has begun to diffract and has waist which is significantly larger than the electron beam.
    c) shows the plasma density profile and the evolution of the laser field amplitude during the simulation.
    d) shows the electron spectrum at \unit[$z=25$]{mm}.
    }
    \label{fig:LWFAsims}
\end{figure}


To explore the dynamics of the LWFA interaction, and to estimate the electron and laser distributions at the collision point, we performed particle-cell-simulations using FBPIC \cite{Lehe2016CPC}.
In the simulation, a laser pulse was initialised to match the central mode of the experimentally measured focal spot with a pulse length of \unit[40]{fs}.
The pulse length was assumed to be larger than the bandwidth limit due to undiagnosed spatiotemporal couplings \cite{Pariente2016NaturePhoton}.
The laser was set to focus \unit[10]{mm} into the plasma and had an initial energy of \unit[5.8]{J}, as the majority of the laser energy was not contained within the central mode.

The simulated plasma density profile was matched to offline neutral gas density measurements (Fig. \ref{fig:LWFAsims}c).
The peak electron density was scaled to match the simulated electron beam properties to those observed in the experiment, achieving best agreement for $n_e = 1.75\times\unit[10^{18}]{cm^{-3}}$.
The 2\% nitrogen dopant was initially ionised up to the 5th level and the helium was fully ionised.
The laser field was able to fully ionise the nitrogen and enable ionization injection as shown in Fig. \ref{fig:LWFAsims}a).

While in the plasma, the laser drives a highly nonlinear plasma wave and experiences self-guiding, maintaining a high $a_0$, throughout the propagation \cite{Lu2007PRSTAB}.
After propagating to $\unit[z=25]{mm}$ (the closest collision point in the experiment), the laser had diffracted to a spot size of \unit[70]{{\textmu}m} \textsc{fwhm} and was trailed by an electron beam with $\unit[1.0]{nC}$ above \unit[260]{MeV} up to a maximum energy of $\unit[1.5]{GeV}$ (Fig. \ref{fig:LWFAsims}b) and d)). 
The laser peak amplitude at this position was $a_0=2.0$ (\ref{fig:LWFAsims}c) with an average wavelength of \unit[1]{{\textmu}m} due to redshifting in the plasma.
For a wavelength of \unit[1]{{\textmu}m}, the Rayleigh length $z_R = \pi w_0^2/\lambda$ fitted to the experimental data in Fig. \ref{fig:ICSYield} corresponds to \unit[$w_0=27$]{{\textmu}m}, consistent with the guided laser spot size in the simulation.
The simulated electron beam had a transverse rms beam size of \unit[4.3]{{\textmu}m}, and a bunch length of \unit[9.6]{\textmu m}.

We can also infer the collision parameters by simulating the laser-electron interaction using the code \textsc{Ptarmigan} \cite{BlackburnPoP2023}, including the experimentally measured electron spectra and varying the laser field amplitude $a_0$ to match the measured photon energies. 
For the measured electron energies and realistic laser intensities ($a_0 < 5$), we expect quantum effects and the impact of radiation reaction to be negligible (quantum nonlinearity parameter $\chi < 0.05$, energy loss $\Delta E <\unit[5]{\%}$ \cite{ThomasPRX2012}).
As a result, the measured post-interaction spectrum is a suitable approximation of the electron beam at the interaction.
For most of the results, the simulated critical energies matched the experimental values for laser fields with $a_0<2$, assuming a wavelength of \unit[1]{{\textmu}m}, in agreement with the PIC simulations.
The only exception is the brightest shot that found a closer match for $a_0 \approx 5$.
We attribute this to fluctuations in the incident laser pulse and also in the plasma density close to the tape, both of which can affect self-guiding of the laser.
Overall, the simulations are in approximate agreement with the experimental measurements, but providing a more definitive estimate of the interaction conditions (as required for SFQED studies \cite{PoderPRX2018,ColePRX2018,MirzaieNatPhoton2024}) is challenging due to the complexity of the interaction and requires additional diagnostics.
This could include, for instance, direct on-shot measurements of the spatial, temporal and spectral properties of the laser pulse after exiting the LWFA and reflecting off the plasma mirror.

\begin{figure} 
    \centering
    \includegraphics[width=0.99\linewidth]{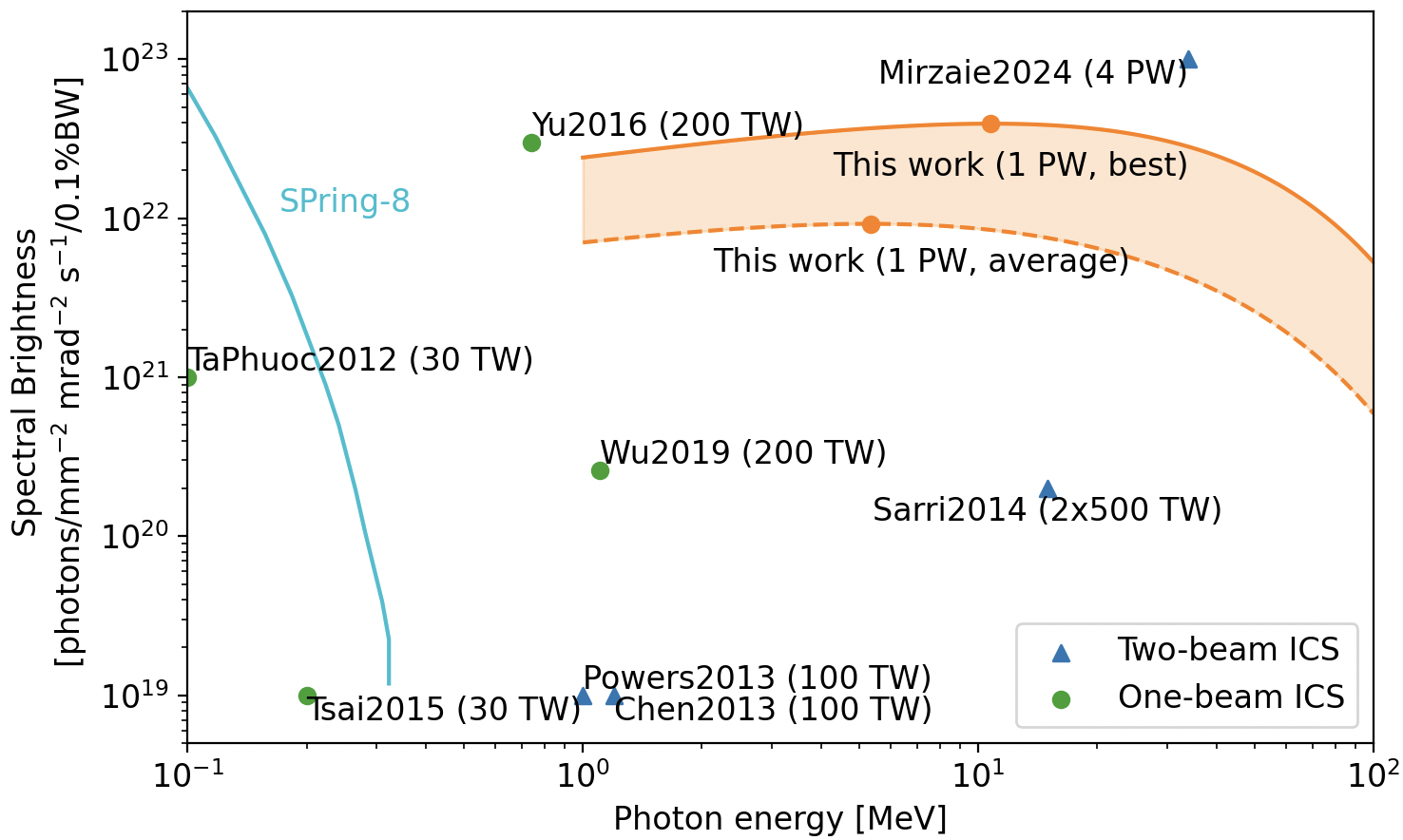}
    \caption{Spectral brightness of our work (orange) compared to other all-optical radiation sources
    using single-beam \cite{TaPhuocNatPhoton2012,TsaiPoP2015,YuSciRep2016,WuPPCF2019} (green circle) or two-beam ICS \cite{ChenPRL2013,PowersNatPhoton2014,SarriPRL2014,KhrennikovPRL2015,MirzaieNatPhoton2024} (blue triangle), and the SPring-8 synchrotron source (extracted from \cite{FletcherNatPhoton2015}). 
    Lines indicate spectral brightness and markers peak spectral brightness.
    Radiofrequency-based ICS sources provide high average flux of order $10^{14}$ photons/s at lower peak brightness $<10^{15}$ \cite{DeitrickPRAB2018,GuntherSynchrotronRad2020} up to MeV photon energies, at an average flux of \unit[$10^{6} - 10^{10}$]{photons/s} at the MeV to GeV scale \cite{OhgakiNIMA2000,KawaseNIMA2008,WellerPPN2009,WangNST2022}
    }
    \label{fig:Brilliance}
\end{figure}

By combining the electron beam distribution from the PIC simulations with the measured properties of the photon source, we can estimate the spectral brightness and compare its performance with other sources (see Fig. \ref{fig:Brilliance}).
The peak brightness, averaged over the 10 shots in Fig. \ref{fig:ICSClose}, was $\unit[(0.9 \pm 0.5) \times 10^{22}]{photons/mm^2/mrad^2/s/0.1\%BW}$ at a photon energy of \unit[5]{MeV}.
The single brightest shot had a peak brightness of $\unit[(3.9 \pm 1.5) \times 10^{22}]{photons/mm^2/mrad^2/s/0.1\%BW}$ at \unit[11]{MeV}. 
This is over 100 times higher than previous two-beam results at a similar photon energy \cite{SarriPRL2014}, and has a similar peak brightness to previous single-beam results \cite{YuSciRep2016} but at 10 times higher photon energy.
The peak brightness is just under half of the value reported from a two-beam experiment by Mirzaie \emph{et al.} \cite{MirzaieNatPhoton2024}, which used about four times the laser power.
However, the photon flux in our case is 100 times higher \edit{and collisions are observed on every shot due to the self-aligning geometry}.
\edit{The high photon yield is enabled by the interaction with a high-charge electron beam ($\unit[\approx 1]{nC}$) and improved scattering efficiency that maximises the participating charge.
In contrast, in the experiment of Mirzaie \textit{et al.} \cite{MirzaieNatPhoton2024}, less than 1\% of electrons were overlapped with the laser pulse to optimise collision probability. 
Importantly, for many practical applications, high photon flux, reduced laser requirements, and consistent operation are critical performance metrics, making this approach particularly attractive.}
The lower peak brightness \edit{observed here is primarily} due to the comparably large divergence of the photon beam, \edit{which is} inherited from \edit{the divergence} of the electron beam (\unit[$\approx 4$]{mrad}).



The nominal performance of a single-beam all-optical ICS can be estimated by considering the scalings for the non-linear self-guided regime, as given by Lu \emph{et al.} \cite{Lu2007PRSTAB}.
In this case the pulse dimensions are $\omega_p w_0/c = \omega_p \tau =  2 \sqrt{a_{g,0}}$, where $\omega_p$ is the plasma frequency. 
The electron beam energy and charge are given by $\gamma m_e c^2 = \frac{2}{3} a_{g,0} m_e c^2 \omega_0^2 / \omega_p^2$ and $e N_b =e (k_p r_b)^3/(30 k_p r_e)$ respectively, where $a_{g,0}$ is the guided laser normalised vector potential and $\omega_0$ is the laser frequency.
Assuming 50\% reflectivity at the plasma mirror (i.e. $a_{0} = a_{g,0}/\sqrt{2}$), and neglecting non-linear evolution of the laser spectrum and pulse length, the radiated energy from the ICS interaction can be estimated from the classical relativistic Larmor formula \cite{LandauLifshitzQED} (valid only for $\chi\ll 1$),
\begin{align}
    W_{ICS} &\approx \frac{2}{3} \frac{e^2 m_e^2 c^3}{4\pi\epsilon_0 \hbar^2} \chi^2 N_b \frac{\tau}{2} \\
    &\approx \left[\frac{1024}{405\pi^2}\frac{ r_e^2}{m_e c^4} \right] \frac{\omega_0^2 W_L^2}{a_{g,0}}
\end{align}
where $W_L$ is the guided laser energy, $\tau/2$ is the interaction time in counterpropagating geometry and $\chi \approx 2   \hbar\omega_0 \gamma a_0/m_e c^2$.
For $a_{g,0}=4$ (nominal value for the nonlinear bubble regime), this reduces to $W_{ICS} \approx  \edit{\eta_{ICS} W_L \unit[]{[J]}}$, where $\eta_{ICS} = 4\times10^{-4} \edit{\unit[]{[J^{-1}}]} W_L$ acts as a \edit{dimensionless} energy-dependent efficiency term.
In the case of our experiment, the amount of laser energy participating in the LWFA was \unit[$W_L\approx5.8$]{J} and so the expected ICS gamma beam energy is \unit[$W_{ICS}\approx13$]{mJ}, in approximate agreement with the brightest experimental measurement\edit{ of $\unit[21\pm9]{mJ}$}.
While ICS beam generation becomes more efficient at higher laser energy, further increasing the efficiency term could be achieved through increasing the laser intensity at the collision, for example, by increasing the plasma density close to the plasma mirror such that the laser self-focuses to a higher $a_0$.

The average photon flux is lower than that achieved in conventional ICS sources \cite{WangNST2022}, but its extremely high peak flux and ultrafast pulse duration provide novel capabilities that are complementary to those of conventional sources.
For instance, they could enable advanced blur-free imaging of high-Z materials, and studies of fundamental physics, e.g. photon-photon scattering \cite{DrebotPRAB2017}.
At a repetition rate of \unit[1]{Hz}, achievable by reducing the gas load on the vacuum system, the source could provide an average photon flux of $\unit[\sim10^9]{photons/s}$ above \unit[1]{MeV} in a broadband spectrum with $\unit[\sim10^6]{photons/0.1\%BW/s}$ at \unit[10]{MeV}.
This photon flux enables the use of nuclear resonance fluorescence \cite{BertozziNIMPB2005}, e.g., to detect clandestine fissile material hidden in cargo or the non-destructive assay of fissile material from spent nuclear fuel \cite{QuiterNIMP2011,Quiter2012PhysRevC}, which could then be transmutated into less hazardous radionuclides \cite{ChenCPC2008,LiJNT2009,RehmanANE2017}.

In conclusion, we have demonstrated a single-beam all-optical Compton source at the PW scale that reaches photon energies \unit[$\epsilon_\gamma>10$]{MeV} at a high photon yield of \unit[$N_\gamma>10^9$]{photons} above \unit[1]{MeV}, improving on previous demonstrations using this scheme at lower laser powers.
A simple model, assuming the LWFA drive laser remains guided up to the plasma mirror, matches the observed ICS beam energy and highlights pathways for increasing the efficiency of the process, for example by tailoring the plasma density or using focusing plasma mirrors \cite{TsaiPoP2015,YuPoP2024} to increase the collision intensity.
Extending this work to higher repetition rates would result in a compact high-average-flux gamma source that could complement radiofrequency-based ICS and bremsstrahlung sources in nuclear physics applications.


\begin{acknowledgments}
We acknowledge funding from the Royal Society URF-R1221874 (M.J.V.S. and E.G.)., the EPSRC grant No. EP/V044397/1 and EP/V049186/1 (G.S.), EP/V049577/1 (B.K. and S.P.D.M.), EP/V049461/1 (C.P.R.), US NSF-GACR collaborative grant 2206059 (A.G.R.T. and B.K.K.), US NSF grant 2108075 and US DOE NNSA Center of Excellence Cooperative Agreement No. DE-NA0003869 (A.G.R.T. and R.F.), and support of the Vulcan dark period community support programme 24-1.
This work was supported by the Extreme Light Infrastructure–Nuclear Physics (ELI-NP) Phase II, a project co-financed by the Romanian Government and the European Union through the European Regional Development Fund, by the Romanian Ministry of Education and Research CNCS-UEFISCDI (Project No. PN-IIIP4-IDPCCF-2016-0164) and Nucleu Projects Grant No. PN 23210105. 
The Romanian Government also supports ELI-NP through IOSIN funds as a Facility of National Interest.
The authors also thank C.A.J. Palmer for valuable discussions.
\end{acknowledgments}

\bibliography{references}

\end{document}